\documentclass[12pt]{iopart}

\usepackage{graphicx}
\usepackage{iopams}

\def\a{\alpha}
\def\b{\beta}
\def\t{\tau}
\def\g{\gamma}
\def\d{\epsilon}
\def\w{\omega}
\def\ket{\rangle}
\def\bra{\langle}
\def\r{\rho}
\def\th{\theta}
\def\s{\sigma}
\def\up{\uparrow}
\def\dn{\downarrow}
\def\ri{\rightarrow}

\def\br{\nonumber \\}

\begin{document}

\title[Synchronisation of kinks in the two-lane TASEP with open
boundaries] {Synchronisation of kinks in the two-lane totally asymmetric
simple exclusion process with open boundary conditions} 

\author{Tetsuya Mitsudo\footnote[1]{mitsudo@scphys.kyoto-u.ac.jp} and Hisao Hayakawa\footnote[2]{hisao@yuragi.jinkan.kyoto-u.ac.jp}}

\address{Department of Physics, Yoshida-south campus, Kyoto University, Sakyo-ku, Kyoto, Japan, 606-8501}

\begin{abstract}
We study the motion of kinks in a
two-lane model of the totally asymmetric simple exclusion
 process with open boundaries.
We analytically study the motion of the kinks by a decoupling approximation. 
In terms of the decoupling approximation, we find that the positions of
 the kinks become synchronised,
though the difference in the number of particles between lanes remains
 non-zero when the rate of lane change is asymmetric.
The validity of this result is confirmed for small asymmetric cases
 through the Monte Carlo simulation.

\end{abstract}

\section{Introduction}

We often encounter congestion of pedestrian flows and traffic flows in
our daily life. 
We also observe the stuck of grains in granular flows.
It is important to study the mechanism of congestion not only from
the industrial point of view but also from the physical point of view.
For the sake of scientific research, we need 
to analyse a simple model which captures the essence of the phenomena.

The asymmetric simple exclusion process (ASEP) is one of the simple models
adequate to describe such the transport phenomena \cite{GS1,SZ}. 
It is a stochastic system of particles moving asymmetrically on a lattice.
The simplest limit of ASEP is that the particle is only allowed to hop
in one direction, which is called the totally asymmetric simple exclusion
process (TASEP).

It is known that the stationary state of one-lane ASEP under open
boundary conditions has been obtained exactly \cite{DEHP,GS3,SS,USW}.
Dynamical properties of TASEP are also studied expensively.
The exact solutions of the master equation by Bethe ansatz on an infinite
system \cite{GS2} and a periodic system \cite{PRE} have been obtained. 
Furthermore, the current fluctuations in an infinite system and a
semi-infinite system \cite{PS} are also studied. 
In the open boundary system, we can draw a phase diagram by the
parameters of inflow rate and outflow rate at the boundaries.
On the phase boundary between the low density phase and the high
density phase, there exists a diffusive domain wall (kink) \cite{DLS,FE,KSKS,SA,TMH}.
Recently, Takesue \etal \cite{TMH} have derived a $f^{-3/2}$ law
in the  power spectrum as a function of the frequency $f$ based on the random walk picture of the kink, and
confirmed its quantitative validity from the comparison of
the Monte Carlo simulation with their theoretical prediction.

However we little know the properties of a multi-lane ASEP which is 
more realistic than the one-lane model.
There are several two-lane models of ASEP \cite{BKNS,Nag,POP,PB}. 
As used in \cite{BKNS} or \cite{Nag}, a realistic lane change rule
should refer to the states of the front sites.
However, the rule makes us difficult to analyse because we need to
construct a transfer matrix to refer to the states of three or four sites,
which are the current site, the front site and the side site, or
the front site of another lane.
Belitsky \etal \cite{BKNS} successfully analysed the
long-time properties of such the two-lane model in an infinite system.
Here, we adopt a simpler model of lane change in which the
particle may change lanes when the side site is vacant and do not refer to the
front site.
Although the two-lane(channel) model which adopts this simple rule is dealt by
Pronina \etal \cite{PB} who analysed the model based on a cluster approximation and compared the result with their 
extensive simulations,
 the lane change rates and the boundary
parameters are symmetric in their model.
We extend  their model to the case of asymmetric lane change rule and boundary parameters to study a
more general situations.  

The purpose of this paper is to clarify the motion of kinks
in a two-lane TASEP.
To fulfil the analysis, we introduce our model and explain how to
specify the position of the kink in the next section.
In section 3, we discuss the motion of two kinks based on a
decoupling (mean field) approximation. 
We find that the motion of the kinks are synchronised though the
number of particles in one lane is different from that in another lane.
We compare the solution with the results of Monte Carlo simulation.
In section 4, we discuss the validity of the mean field approximation.
We find that the two-point correlation function is small during the relaxation
process from the independent motion of two kinks to a synchronised
motion of them.
In section 5, we conclude our results.

\section{Our model}

\subsection{Introduction of our two-lane model}

Our two-lane model is defined on a two lane lattice of $L\times 2$
sites, where $L$ is the length of one lane.
We introduce the occupation variable $\t_{j;\ell}$ where $\t_{j;\ell}=1$
and $\t_{j;\ell}=0$ represent the occupied state and the vacant state on
the $j$th site in the $\ell$th lane, respectively.
The particle move forward  by the rate $1$ during the time interval
$\rmd t$ if the front site is vacant.
We assume that all the particles drift from the left to the right.
The open boundary condition is characterised by the inflow rate
$\a_{\ell}$ and the outflow rate $\b_{\ell}$.
The particle is injected to the system by
the rate $\a_{\ell}$ when the first site in the $\ell$th lane is empty,
while the particle is extracted from the system by the rate $\b_{\ell}$ 
when the $L$th site in the $\ell$th lane is occupied.
On all sites, the particle is allowed to change lanes only to the
neighbouring site.
A particle on the $1$st($2$nd) lane can change lanes by the rate
$r_{\dn}$($r_{\up}$) when the side site in another lane is empty.

\begin{figure}
\begin{center}
\includegraphics[scale=1.0]{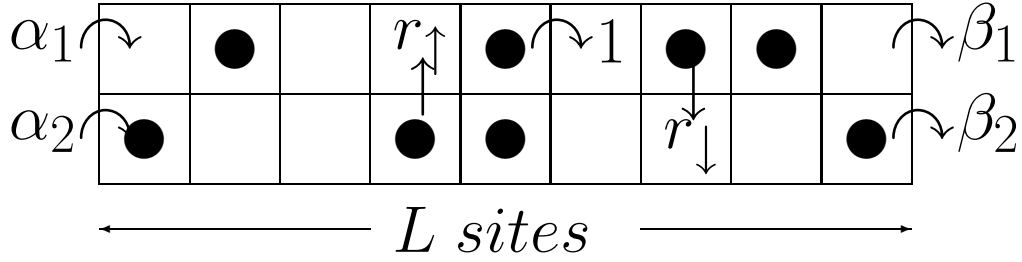}
\end{center}
\caption{A sample picture of the two-lane model.}
\end{figure}

We denote the probability of finding the system in the configuration
$\btau=\{\t_{1;1},\t_{2;1},\cdots,\t_{L;2}\}$ by
$P(\t_{1;1},\t_{2;1},\cdots,\t_{L;2})$.
We write the time evolution of the two-lane model by the master
equation,
\begin{eqnarray}
\fl \frac{\rmd}{\rmd t}P(\t_{1;1},\cdots,\t_{L;2})=\sum_{\s_{1;1}}(h_{1;1})_{\t_{1;1};\s_{1;1}}P(\s_{1;1},\cdots,\t_{L;2})+\sum_{\s_{1;2}}(h_{1;2})_{\t_{1;2};\s_{1;2}}P(\t_{1;1},\cdots,\s_{1;2},\cdots,\t_{L;2}) \br
\lo +\sum_{\ell=1}^2\sum_{j=1}^{L-1}\sum_{\s_{j;\ell},\s_{j+1;\ell}}(h_{j,j+1;\ell})_{\t_{j;\ell},\t_{j+1;\ell};\s_{j;\ell},\s_{j+1;\ell}}P(\t_{1;1},\cdots,\s_{j;\ell},\s_{j+1;\ell},\cdots,\t_{L;2}) \br
\lo +\sum_{\s_{L;1}}(h_{L;1})_{\t_{L;1};\s_{L;1}}P(\t_{1;1},\cdots,\s_{L;1},\cdots,\t_{L;2}) 
+\sum_{\s_{L;2}}(h_{L;2})_{\t_{L;2};\s_{L;2}}P(\t_{1;2},\cdots,\s_{L;2}) \br
+\sum_{j=1}^L\sum_{\s_{j;1},\s_{j;2}}(h_{j;1,2})_{\s_{j;1},\s_{j;2};\t_{j;1},\t_{j;2}}P(\t_{1;1},\cdots,\s_{j;1},\cdots,\s_{j;2},\cdots,\t_{L;2})
\end{eqnarray}
where $\s_{j;\ell}$ is used for a dummy variable in the summation, 
and the transition matrices $h_{1;\ell},h_{L;\ell},h_{j,j+1;\ell},h_{j;1,2}$
are represented as
\begin{eqnarray}
\eqalign{
h_{1,\ell}=\left(\begin{array}{cc} -\a_{\ell} & 0 \\ \a_{\ell} & 0 \end{array}\right)_{1;\ell} \qquad
h_{L,\ell}=\left(\begin{array}{cc} 0 & \b_{\ell} \\ 0  & -\b_{\ell} \end{array}\right)_{L;\ell} \\
h_{j,j+1;\ell}=\left(\begin{array}{cccc}
0 & 0 & 0 & 0 \\
0 & 0 & 1 & 0 \\
0 & 0 & -1 & 0 \\
0 & 0 & 0 & 0
\end{array}\right)_{j,j+1;\ell} \qquad
h_{j;1,2}=\left(\begin{array}{cccc}
 0 & 0 & 0 & 0 \\ 0 & -r_{\up} & r_{\dn} & 0 \\ 0
	  & r_{\up} & -r_{\dn} & 0 \\ 0 & 0 & 0 & 0 \end{array}\right)_{j;1,2}.}
\end{eqnarray}
Here the density function $\bra \t_{j;\ell}\ket$ and two-point
function $\bra \t_{j;\ell} \t_{k;\ell'}\ket$ are defined by,
\begin{eqnarray}
\bra \t_{j;\ell}\ket = \sum_{\btau}\t_{j;\ell}P(\t_{1;1},\cdots,\t_{L;2}) \\
\bra \t_{j;\ell}\t_{k;\ell'}\ket=\sum_{\btau}\t_{j;\ell}\t_{k;\ell'}P(\t_{1;1},\cdots,\t_{L;2}),
\end{eqnarray}
where the summation is taken over all the configurations.
The time evolution of $\bra \t_{j;\ell}\ket$ is written as
\begin{eqnarray}
\label{avril}
\frac{\rmd}{\rmd t}\bra \t_{j;\ell} \ket =
 J_{j-1,j;\ell}-J_{j,j+1;\ell}-J_{j;\ell\ri \ell'}+J_{j;\ell'\ri \ell}
\end{eqnarray}
for $\ell'\neq\ell$,
where the current $J_{j,j+1;\ell}$ between site $j$ and $j+1$ is 
\begin{equation}
J_{j,j+1;\ell}=\bra\t_{j;\ell}(1-\t_{j+1;\ell})\ket 
\end{equation}
and the currents between lanes are 
\begin{eqnarray}
J_{j;1\ri 2}=r_{\dn}\bra \t_{j;1}(1-\t_{j;2}) \ket \qquad
J_{j;2\ri 1}=r_{\up}\bra  \t_{j;2}(1-\t_{j;1})\ket.  
\end{eqnarray}

\subsection{The Position of the Kink}

It is known that a kink appears when the inflow rate is equal to
 the outflow rate and both rates are smaller than $1/2$ in one lane ASEP.
The kinks also appear in the two-lane model when
$\a_1=\b_1<1/2$ and $\a_2=\b_2<1/2$. 
For $r_{\up}\neq 0$ and $r_{\dn}\neq 0$, it is obvious that the 
motion of one kink depends on another kink. 

We need to specify the position of the kinks in order to discuss their
correlated motion. 
The position of a stable kink in the one-lane ASEP can be determined by using the
second class particle \cite{DLS,FE}.
However, we adopt another method to determine the position of the kink by
the whole number of particles in a lane.
This definition has been used in the domain wall theory \cite{KSKS,SA,TMH},
 and give the exact position of the kink when the inflow and
 outflow rates are small. 
The advantage to adopt this method is that it is much simpler than the
method by the second class particle. 

We introduce $\bra N_{\ell}\ket$ for the whole number of particles in each lane
\begin{eqnarray}
\bra N_{\ell}\ket=\sum_{j=1}^L \bra \t_{j;\ell} \ket .
\end{eqnarray}
We also introduce $\bra N_G\ket$ and $\bra N_R\ket$ by
 $\bra N_G\ket=(\bra N_2\ket + \bra N_1 \ket)/2$ 
 and $\bra N_R\ket=\bra N_2\ket - \bra N_1\ket$ respectively.
The position of the kink $x_{\ell}$ is defined from the equation based
 on a kink picture;
\begin{equation}
x_{\ell}=\frac{\bra N_{\ell}\ket-\r_{\ell;+}L}{\r_{\ell;-}-\r_{\ell;+}} ,\label{n1}
\end{equation}
where $\r_{\ell;\pm}$ represent the density of the $\ell$th lane.
The index $+$ represents the right side of the position of the kink
and index $-$ represents the left side of the position of the kink.
It is straightforward to give the eq.(\ref{n1}) by summing up the 
equation 
\begin{equation}
\bra \t_{j;\ell} \ket=\r_{\ell;-}+(\r_{\ell;+}-\r_{\ell;-})\th(j-x_{\ell}) \label{kin}
\end{equation}
from $j=1$ to $L$, where $\th(z)$ is the step function,
\begin{equation}
\th(z)=\cases{
1 & for $z\geq 0$ \\
0 & for $z<0$ \\}. 
\end{equation}
Thus, once $\bra N_{\ell}\ket$ is known, we can determine the position
of the kink.

\section{Mean-Field Theory}

For the large system size $L$, we can neglect the
boundary terms. 
Thus the equations for $\bra N_G\ket,\bra N_R\ket$ are given by
\begin{eqnarray}
\frac{\rmd}{\rmd t}\bra N_G \ket &=& 0 \\
\frac{\rmd}{\rmd t}\bra N_R \ket &=& -2r_{\up}\bra N_R \ket +2(r_{\dn}-r_{\up})\sum_{j=1}^L \bra \t_{j;1}(1-\t_{j;2}) \ket. \label{for} 
\end{eqnarray}
When $r_{\dn}=r_{\up}$, the eq.(\ref{for}) is reduced to
\begin{equation}
\frac{\rmd}{\rmd t}\bra N_R \ket = -2r_{\up}\bra N_R \ket.
\end{equation}
Thus $\bra N_R\ket$ relaxes to $0$ exponentially,
and the number of particles becomes identical in both lanes in the long
time limit.

However for $r_{\dn}\neq r_{\up}$, the problem becomes nontrivial
because of the two-point function in the second term in the right hand
side of the eq.(\ref{for}). 
In general, the two-point correlation function is determined by an equation
including three-point correlation function.
Thus, we cannot obtain the exact form of the many-point correlation functions 
without truncation of the hierarchy of correlation functions.
Here we adopt the simplest truncation, which is the decoupling
(mean-field) approximation as 
\begin{equation}
\sum_{j=1}^L \bra \t_{j;1}(1-\t_{j;2}) \ket \simeq \sum_{j=1}^L \bra \t_{j;1}\ket(1-\bra \t_{j;2} \ket) . \label{apr}
\end{equation}
We also use the kink picture (\ref{kin}) to approximate the
density profile $\bra\t_{j;\ell}\ket$.

Let us discuss the motion of two kinks starting from the initial
condition where two separated kinks exist in both lanes.
The density profile changes in time during the synchronisation of the kinks.
Furthermore, we assume that the density changes by keeping the density
profile (\ref{kin}).
Thus we have to determine the time evolution of the density
$\r_{\ell;\pm}$ on the both sides of the kink.
By introducing  $\r_{G;\pm} =\frac{\r_{2;\pm}+\r_{1;\pm}}{2}$ 
and $\r_{R;\pm} = \r_{2;\pm}- \r_{1;\pm}$,
the time evolution equations for $\r_{G;\pm}$ and $\r_{R;\pm}$ are
respectively written as
\begin{eqnarray}
 \frac{\rmd}{\rmd t}\r_{G;\pm} =0 \label{me}\\
\frac{\rmd}{\rmd t} \r_{R;\pm} 
= -(r_{\up}+r_{\dn}) \r_{R;\pm} + \frac{1}{2}(r_{\dn}-r_{\up})
 \r_{R;\pm}^2+(r_{\dn}-r_{\up})(
 \r_{G;\pm}-\r_{G;\pm}^2) \label {my}.
\end{eqnarray}
Equations (\ref{me}) and (\ref{my}) can be solved exactly,
\begin{eqnarray}
\r_{G;\pm} =\r_{G;\pm}^0 \label{const}\\
\r_{R;\pm} = \w_- + \frac{2\g}{\d}\frac{e^{-\g t}}{e^{-\g t}-C_{\pm}},
\end{eqnarray}
where
\begin{eqnarray}
\d=r_{\dn}-r_{\up} \qquad
\g=\frac{1}{2}(r_{\dn}-r_{\up})(\w_+ - \w_-) \\
\w_{\pm} = \frac{r_{\dn}+r_{\up}\pm\sqrt{(r_{\dn}+r_{\up})^2-2(r_{\dn}-r_{\up})^2(\a_1+\a_2- (\a_1+\a_2)^2/2)}}{r_{\dn}-r_{\up}} \\
C_{\pm}=\frac{\r_{R;\pm}^0-\w_+}{\r_{R;\pm}^0-\w_-}.
\end{eqnarray}
The initial conditions $\r_{G;\pm}^0$ and $\r_{R;\pm}^0$ are taken as
the stationary densities in one-lane model as
\begin{eqnarray}
\r_{G;-}^0 =\frac{\a_1+\a_2}{2} \qquad \r_{R;-}^0=\a_2-\a_1 \label{l1} \\
\r_{G;+}^0 =\frac{2-\a_1-\a_2}{2} \qquad \r_{R;+}^0=\a_1-\a_2 \label{l2}.
\end{eqnarray}

Thus we obtain the density $\r_{\ell;-}$
\begin{eqnarray}
\r_{1;-}=\r_1'-\frac{\g}{\d}\frac{e^{-\g t}}{e^{-\g t}-C_{-}} \qquad
\r_{2;-}=\r_2'+\frac{\g}{\d}\frac{e^{-\g t}}{e^{-\g t}-C_{-}} 
\end{eqnarray}
and the density $\r_{\ell;+}$
\begin{eqnarray}
\r_{1;+}=1-\r_2'-\frac{\g}{\d}\frac{e^{-\g t}}{e^{-\g t}-C_{+}} \qquad
\r_{2;+}=1-\r_1'+\frac{\g}{\d}\frac{e^{-\g t}}{e^{-\g t}-C_{+}} 
\end{eqnarray}
where
\begin{eqnarray}
\r_1'=\frac{\a_1+\a_2-\w_-}{2} \qquad
\r_2'=\frac{\a_1+\a_2+\w_-}{2} 
\end{eqnarray}
Therefore we obtain the time evolution of the density profile $\bra
\t_{j;\ell}\ket$ for $r_{\dn}\neq r_{\up}$.

Substituting eq.(\ref{kin}) into (\ref{for}) with the aid of
(\ref{apr}) we obtain,
\begin{eqnarray}
\fl
\frac{\rmd}{\rmd t}\bra N_R \ket 
= -(2r_{\up}+\d(1-\r_{2;-}+\r_{1;+}))\bra N_R \ket +2\d(\r_{1;+}\r_{2;-}L+(1-\r_{1;+}-\r_{2;-})\bra N_G \ket)  \label{asol}
\end{eqnarray}
for $x_1<x_2$ and
\begin{eqnarray}
\fl
\frac{\rmd}{\rmd t}\bra N_R \ket 
= -(2r_{\up}+\d(1-\r_{2;+}+\r_{1;-}))\bra N_R \ket +2\d(\r_{1;-}\r_{2;+}L+(1-\r_{1;-}-\r_{2;+})\bra N_G \ket) \label{sol} 
\end{eqnarray}
for $x_1>x_2$.
Equations (\ref{asol}) and (\ref{sol}) can be solved exactly, though the
expression is lengthy (see (\ref{A4}) and (\ref{A5})).
Here, we present the solution for $\bra N_R\ket$ in the long time limit as,
\begin{equation}
\bra N_R \ket_{\infty}=\frac{\d\r_1'(1-\r_1')L}{r_{\up}+\d\r_1'} \label{f1}
\end{equation}
for $x_1>x_2$, and 
\begin{equation}
\bra N_R \ket_{\infty}=\frac{\d\r_2'(1-\r_2')L}{r_{\dn}-\d\r_2'}\label{f2}
\end{equation}
for $x_1<x_2$.
These results show that there remains the mean difference of the number of
particles between lanes.
The validity of our analysis based on the decoupling approximation is
confirmed by the comparison of our result with the Monte Carlo
simulation when $|r_{\up}-r_{\dn}|$ is not large.
Figure 2 shows the quantitative accuracy of our analysis in the
time evolution of $\bra N_R\ket$.

\begin{figure}
\label{simu}
\begin{center}
\includegraphics[scale=0.4]{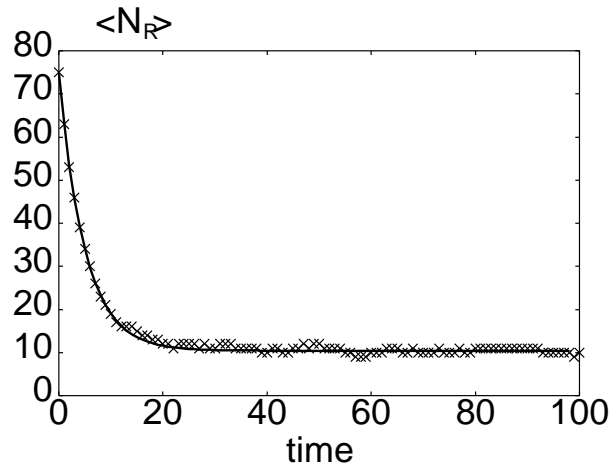}
\caption{The figure shows the comparison of time evolution between
 the simulation and the calculation. The solution (\ref{A4}) is shown by
 the solid line and the simulation  result is shown by the cross($\times$). The parameters are $\a_1=\b_1=0.1,\a_2=\b_2=0.15,r_{\dn}=0.11,r_{\up}=0.1,L=1000$. Time step is taken for each Monte Carlo step. The initial condition is fixed to $\bra N_R \ket_0=75$ and averaged over 1000 samples.}
\end{center}
\end{figure}

Though $\bra N_R\ket$ remains finite, the positions of the kinks are
synchronised. 
In fact, from eqs. (\ref{f1}) or (\ref{f2}) and (\ref{n1}) we obtain,
\begin{equation}
x_2-x_1=0 .
\end{equation}
Thus the positions of the kinks become identical in the long time limit.
This result is reasonable, because we cannot choose a preferable
congestion front in traffic jams.

\section{Discussion}

Now let us discuss the validity of the decoupling approximation.
Although it is difficult to evaluate the two-point function exactly, it
is possible to evaluate it from the Monte Carlo simulation. 
The result of our simulation for the two-point function 
\begin{equation}
A = \frac{\sum_{j=1}^L\bra \t_{j;1}\t_{j;2}\ket-\sum_{j=1}^L\bra \t_{j;1}\ket\bra\t_{j;2}\ket}{\sum_{\ell=1}^2\sum_{j=1}^L\bra \t_{j;\ell}^2\ket} \label{norm}
\end{equation}
is shown in Fig.3 for the same parameters as used in Fig.2, where we realise that $A$ is small between $t=30$ and $t=100$.
The synchronisation is realised before $t=30$ as we can see in the Fig.2. 
Thus, we may expect that the decoupling
approximation adopted here works well to describe the synchronisation of
the kinks. 

\begin{figure}
\begin{center}
\includegraphics[scale=0.4]{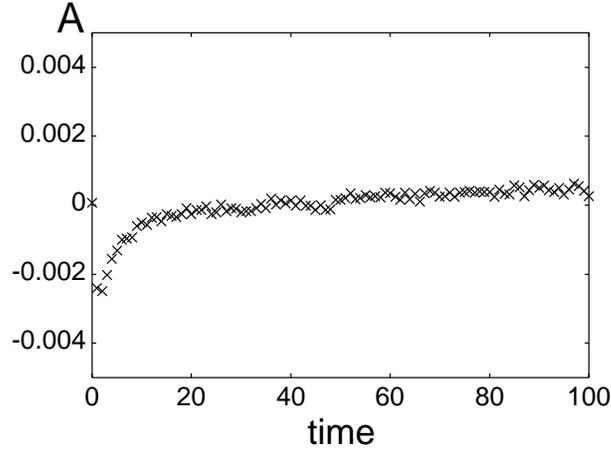}
\caption{ The figure shows the time evolution of the value $A$ from the
 simulation. The parameters used in this simulation are the same as
 those used in Fig.2.}
\end{center}
\end{figure}

\begin{figure}
\label{nor}
\begin{center}
\includegraphics[scale=0.35]{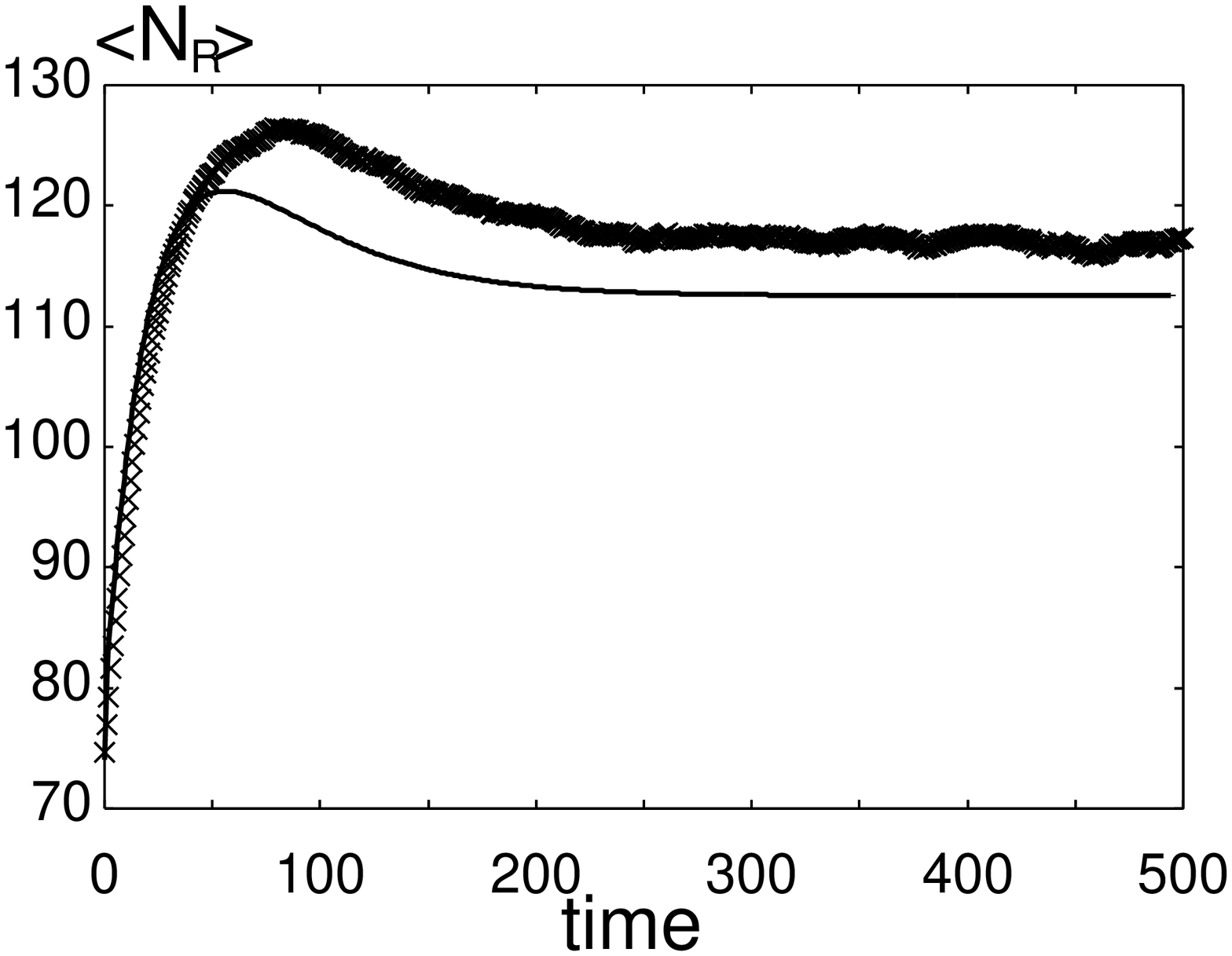}
\includegraphics[scale=0.35]{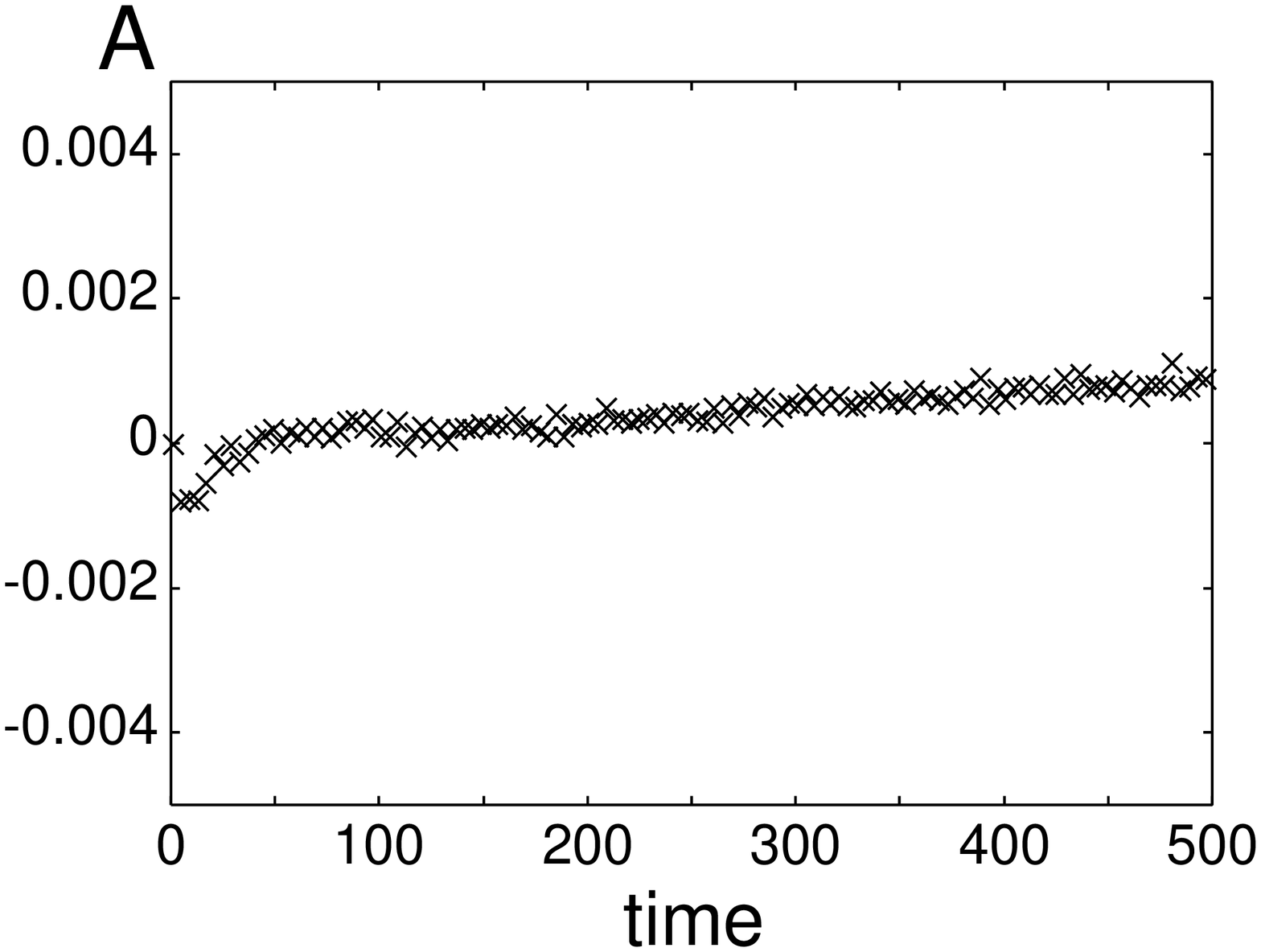}
\caption{The left figure shows the comparison of time evolution between
 the simulation and the calculation. The solution (\ref{A4}) is shown by
 the solid line and the simulation  result is shown by the cross($\times$). The right
 figure shows the time evolution of the value $A$ from the
 simulation. In both figures, the parameters are $\a_1=\b_1=0.1,\a_2=\b_2=0.15,r_{\dn}=0.03,r_{\up}=0.01,L=1000$. Time step is taken for each Monte Carlo step. The initial condition is fixed to $\bra N_R \ket_0=75$ and averaged over 1000 samples.}
\end{center}
\end{figure}

\begin{figure}
\begin{center}
\includegraphics[scale=0.4]{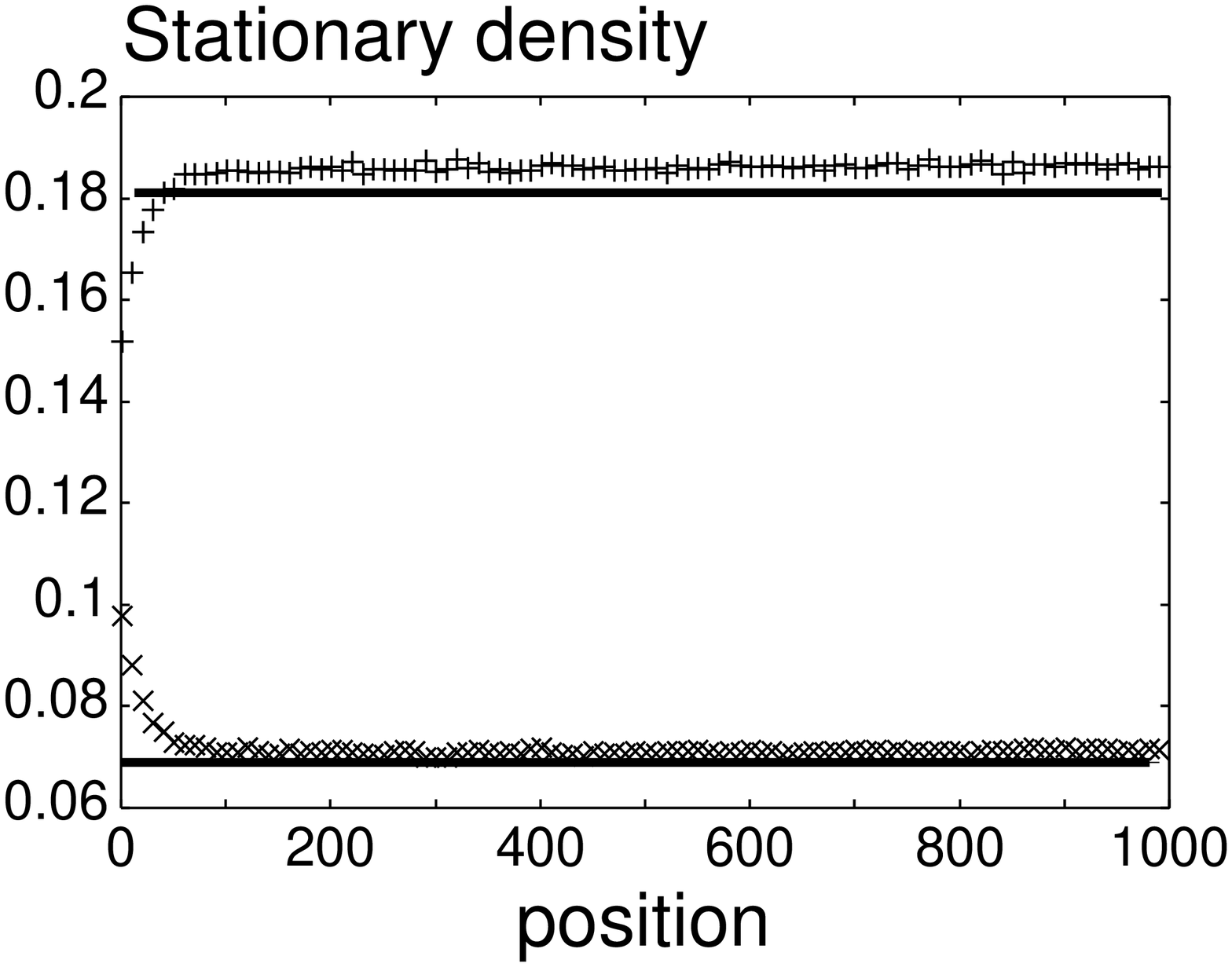}
\caption{The figure shows the difference between the stationary density
 given by the simulation and that by the eq.(\ref{my}), in the
 corresponding low density region. The plus($+$) stands for the first
 lane, the cross($\times$) stands for the density in the second lane and
 the approximation is shown by the solid line.
 The boundary parameters are $\a_1=0.1,\b_1=0.9,\a_2=0.15,\b_2=0.85,r_{\dn}=0.03,r_{\up}=0.01,L=1000$.}
\label{xxx}
\end{center}
\end{figure}

However there is a certain parameter region that the decoupling
approximation fails.
In fact, the left figure of Fig.4 shows $\bra N_R \ket$ obtained
from the simulation deviates from that in the decoupling approximation.
The positions of the kinks are not identical in this case.
The value $A$ given by the simulation in the right figure of Fig.4 when it
deviates from the decoupling approximation.
The value $A$ is almost $0$ in the region between $t=30$
and $t=200$. 
In such the region the stationary density $\r_{\ell;\pm}$ is not
given by the eq.(\ref{my}).
The comparison of $\r_{\ell;-}$ between the result of the simulation and the
solution of the eq.(\ref{my}) is shown in Fig.5.
We can see that the density derived from the mean-field theory 
deviates from the result of the simulation.
Thus in the region that the final densities of each lane differ from the
solutions of the eq.(\ref{my}), the final positions of the kinks  are
not identical. 

\begin{figure}
\begin{center}
\includegraphics[scale=0.4]{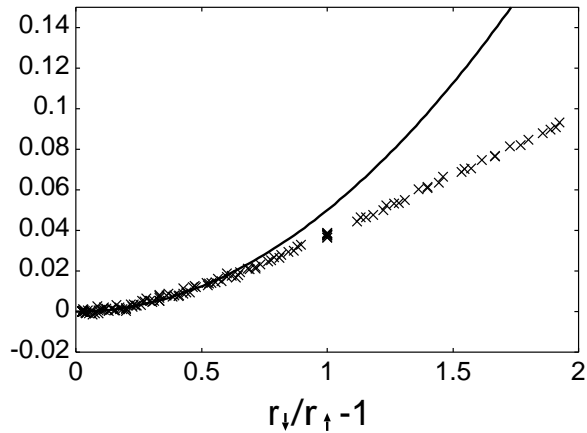}
\caption{The figure shows the final difference between the kink positions
 given by the simulation and the analysis.
The cross($\times$) shows the difference and
 the solid line is the guideline which is proportional to $(\frac{r_{\dn}}{r_{\up}}-1)^2$.
 The boundary parameters are $\a_1=\b_1=0.1,\a_2=\b_2=0.15,L=1000$.}
\end{center}
\end{figure}

To verify the valid region of the decoupling approximation, in Fig.6, 
 we plot the deviation from the decoupling approximation in the
 difference between two kinks 
as a function of $r_{\dn}/r_{\up}-1$. 
From Fig.6 we find that 
the deviation is proportional to $\delta^2=(r_{\dn}/r_{\up}-1)^2$ for the small asymmetric cases.
This is an interesting result, because the decoupling approximation
predicts $\bra N_R \ket\propto \delta$.
Namely, the decoupling approximation can be used when we can regard $\delta$ as finite but $\delta^2$ as negligible.
In addition, the curvature of the parabola is relatively small, which ensures that we may use the approximation in
$\delta< 0.2$.
The quantitative analysis of the violation of the decoupling approximation will be a future problem.

\section{Conclusion}

In this paper, we have studied the motion of kinks in the two-lane TASEP.
We obtain the explicit time evolution function of the average number
of particles in each lane which is related to the position of the kink by
adopting the decoupling approximation of the two-point correlation function.
We find that the positions of the kinks are synchronised, though the
number of particles in a lane can be different from that in another lane. 
We confirm the validity of our analysis by comparing the result of the
Monte Carlo simulation and the analytical result.
The deviation from the mean-field analysis is small when the lane change
rates are nearly symmetric.

We would like to thank S.Takesue for fruitful discussion.
This work is partially supported by the Grant-in-Aid for Scientific
Research (Grant No. 15540393) of the Ministry of
Education, Culture, Sports, Science and Technology(MEXT), Japan, 
and the Grant-in-Aid for the 21st century COE `Center for Diversity and
Universality in Physics' from MEXT, Japan.

\appendix

\section{The calculation of $\bra N_R \ket$}

In this appendix, we give the explicit expression of $\bra N_R\ket$.
Equation (\ref{sol}) is solved as 
\begin{eqnarray}
\fl
\bra N_R\ket &=& \bra N_R\ket_0
 e^{-\int_0^t\rmd t'(2r_{\up}+\d(1-\r_{2;+}+\r_{1;-}))} \br
\fl
& & +\d 
\int_0^t \rmd t' e^{\int_t^{t'} \rmd t''(2r_{\up}+\d(1-\r_{2;+}+\r_{1;-}))}
 (2L\r_{1;-}\r_{2;+}+2(1-\r_{1;-}-\r_{2;+})\bra N_G \ket) \label{label}
\end{eqnarray}
Here, we perform the integral in the argument of the exponential function,
\begin{eqnarray}
 \int_0^t\rmd t'(1-\r_{2;+}+\r_{1;-}) 
= 2\r_1't+\frac{1}{\d}\ln\left(\frac{(C_+ -e^{-\g t})(C_- -e^{-\g t})}{C_+C_-}\right).
\end{eqnarray}
Therefore,
\begin{equation}
e^{-\int_0^t\rmd t'(2r_{\up}+\d(1-\r_{2;+}+\r_{1;-}))}=e^{-2(r_{\up}+\d\r_1')t}\frac{C_+C_-}{(C_+ -e^{-\g t})(C_- -e^{-\g t})}.
\end{equation}
After executing the calculation, we finally achieve
\begin{eqnarray}
\label{A4}
\fl
\bra N_R\ket = \bra N_R\ket_0
e^{-2(r_{\up}+\d\r_1')t}\frac{C_+C_-}{(C_+ -e^{-\g t})(C_- -e^{-\g t})} \br
\lo +\frac{\d}{(C_+ -e^{-\g t})(C_- -e^{-\g t})}\left[L\r_1'(1-\r_1')C_+C_-\frac{1-e^{-2(r_{\up}+\d\r_1')}}{r_{\up}+\d\r_1'}\right. \br
 +\frac{e^{-\g t}-e^{-2(r_{\up}+\d\r_1')}}{2(r_{\up}+\d\r_1')-\g}2L\left(\frac{\g}{\d}(C_+ -\r_1'(C_- +C_+))-(C_- +C_+)\r_1'(1-\r_1')\right) \br
+\frac{e^{-\g t}-e^{-2(r_{\up}+\d\r_1')}}{2(r_{\up}+\d\r_1')-\g}\frac{2\g}{\d}(C_- -C_+)\bra N_G\ket \br
 \left. +\frac{e^{-2\g t}-e^{-2(r_{\up}+\d\r_1')}}{r_{\up}+\d\r_1'-\g}\left(-\frac{\g^2}{\d^2}+\frac{\g}{\d}(2\r_1'-1)+\r_1'(1-\r_1')\right)L \right] 
\end{eqnarray}
for $x_1>x_2$, and 
\begin{eqnarray}
\label{A5}
\fl
\bra N_R\ket = \bra N_R\ket_0
e^{-2(r_{\dn}-\d\r_2')t}\frac{C_+C_-}{(C_+ -e^{-\g t})(C_- -e^{-\g t})} \br
\lo +\frac{\d}{(C_+ -e^{-\g t})(C_- -e^{-\g t})}\left[L\r_2'(1-\r_2')C_-C_+\frac{1-e^{-2(r_{\dn}-\d\r_2')}}{r_{\dn}-\d\r_2'}\right. \br
 +\frac{e^{-\g t}-e^{-2(r_{\dn}-\d\r_2')}}{2(r_{\dn}-\d\r_2')-\g}2L\left(\frac{\g}{\d}(C_+ +\r_2'(C_- -C_+))-(C_- +C_+)\r_2'(1-\r_2')\right) \br
+\frac{e^{-\g t}-e^{-2(r_{\dn}-\d\r_2')}}{2(r_{\dn}-\d\r_2')-\g}\frac{2\g}{\d}(C_- -C_+)\bra N_G\ket \br
 \left. +\frac{e^{-2\g t}-e^{-2(r_{\dn}-\d\r_2')}}{r_{\dn}-\d\r_2'-\g}\left(-\frac{\g^2}{\d^2}+\frac{\g}{\d}(2\r_2'-1)+\r_2'(1-\r_2')\right)L \right]
\end{eqnarray}
for $x_1<x_2$.

\end{document}